\documentclass{Interspeech2024}
\usepackage{multirow}

\interspeechcameraready

\title{Frequency-mix Knowledge Distillation for Fake Speech Detection}

\name{Cunhang}{Fan$^{1}$}
\name{Shunbo}{Dong$^{1}$}
\name{Jun}{Xue$^{1}$}
\name{Yujie}{Chen$^{1}$}
\name{Jiangyan}{Yi$^{2}$}
\name{Zhao}{Lv$^{1}$}

\address{
  $^1$School of Computer Science and Technology, Anhui University, China\\
  $^2$Institute of Automation, Chinese Academy of Sciences, China}
\email{e22201030@stu.ahu.edu.cn, e22201148@stu.ahu.edu.cn}


\begin{document}

\maketitle

\begin{abstract}

In telephony scenarios, the fake speech detection (FSD) task to combat speech spoofing attacks is challenging. Data augmentation (DA) methods are considered effective means to address the FSD task in telephony scenarios, typically divided into time domain and frequency domain stages. While each has its advantages, both can result in information loss. To tackle this issue, we propose a novel DA method, Frequency-mix (Freqmix), and introduce the Freqmix knowledge distillation (FKD) to enhance model information extraction and generalization abilities. Specifically, we use Freqmix-enhanced data as input for the teacher model, while the student model's input undergoes time-domain DA method. We use a multi-level feature distillation approach to restore information and improve the model's generalization capabilities. Our approach achieves state-of-the-art results on ASVspoof 2021 LA dataset, showing a 31\% improvement over baseline and performs competitively on ASVspoof 2021 DF dataset.

\end{abstract}

\renewcommand{\thefootnote}{\fnsymbol{footnote}}
\setcounter{footnote}{0}
\renewcommand{\thefootnote}{\arabic{footnote}}

\section{Introduction}

With the rapid development of text-to-speech (TTS) and voice conversion (VC), it has become increasingly challenging for the human ear to distinguish genuine speech from fake speech. Fake speech detection (FSD) task aims to devise effective counter measures (CM) that bolster the resilience of Automatic Speaker Verification (ASV) system against deceptive attacks.  While previous studies \cite{arif2021voice, zhang2021one, das2019long, nautsch2021asvspoof} have mainly focused on FSD task under controlled laboratory settings, ignoring the practical implications introduced by data encoding, compression, and transmission. FSD task in the logical access (LA) scenario of the ASVspoof 2021 challenge \cite{delgado2021asvspoof} involves training models in controlled laboratory conditions and subsequently deploying them for anti-spoofing tasks on real-world speech data. Addressing the generalization of the model becomes particularly important.

\begin{figure*}[th]
  \centering
  \includegraphics[width=\linewidth]{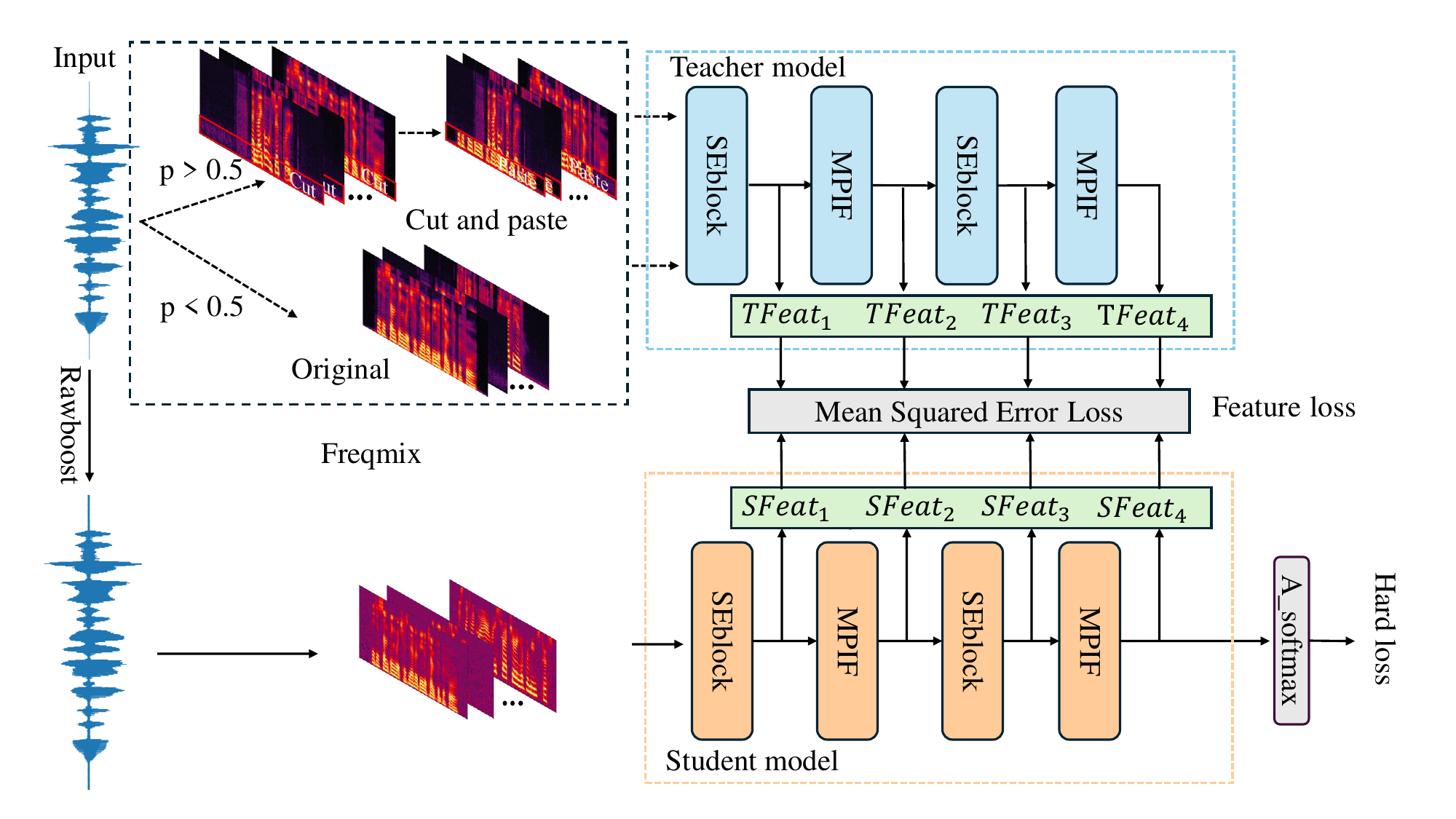}
  \caption{The illustration of our proposed Freqmix knowledge distillation (FKD) for FSD method in telephony scenarios. The student model and the teacher model both adopt the MPIF-Res2Net architecture an identical number of parameters. During the training of the student model, the parameters of the teacher model remain unchanged.}
  \label{fig:speech_production}
\end{figure*}

Data augmentation (DA) consistently stands out as an important technique for improving model generalization. Notably, Rawboost \cite{tak2022rawboost} is devised for the FSD task in scenarios involving communication-distorted and compressed-coded speech data. It operates directly on the raw waveform, causing signal distortion in training samples, thereby enhancing the model's generalization capability. SpecAugment \cite{park19e_interspeech} treats the log mel spectrogram as an image, applying masking on contiguous time steps and mel frequency segments to enhance model performance. Specmix \cite{kim21c_interspeech} combines two training samples to create a new sample, with the labels of the combined corresponding samples serving as the label for the new sample, directly impacting features in the time-frequency domain. MixSpeech \cite{9414483} combines two distinct speech features through weighted blending to generate a novel feature for Automatic Speech Recognition (ASR). While these masking methods can indeed enhance the model's generalization ability, applying them directly to the input of a classification model may potentially erase artefacts directly.

In recent years, the knowledge distillation (KD) method has often been utilized for model compression \cite{hinton2015distilling, fan2024progressive}, condensing a network with thousands of layers into a smaller model. OCKD \cite{lu2023one} combines one-class method with KD, reducing the number of layers in Wav2vec 2.0 \cite{wang22_odyssey, tak22_odyssey} to prevent overfitting caused by an excessive number of parameters. DKDSSD \cite{fan2024dual} is a dual-branch method using KD to address the issue of performance degradation of detecting spoofed speech in noisy environments. However, as the research progresses, researchers have discovered that a student model with the same amount of training parameters as the teacher model can obtain better results than the teacher model \cite{furlanello2018born, zhang2018deep}. \cite{xue2023learning} proposed a self-distillation method, using the deep network as the teacher model and the shallow network as the student model, using deep information to guide shallow information learning to improve fine-grained information recognition. While these works involve KD, we believe that transferring knowledge between completely identical input data may not extract sufficiently rich information.

In this study, we propose a method called Freqmix knowledge distillation (FKD) to improve the information extraction and generalisation capabilities of the model. Taking advantage of the successful performance of Rawboost for the FSD task in telephony scenarios, the input data of our student model is augmented by Rawboost. Freqmix divides the teacher model input into two parts: one with original audio and the other with spectrogram-masked speech samples. The diverse inputs in the teacher model guide the student model to learn varied knowledge. The original data helps to restore artefacts from distorted signals, thus improving the student model information extraction ability. Meanwhile, spectrogram-enhanced data enables the student model to combine the effects of spectrogram masking and signal distortion, thereby boosting the model's generalization capabilities. In conclusion, for the FSD task in telephony scenarios, our proposed FKD method innovatively combines knowledge distillation with different DA methods and original information, further improving the model's performance. Compared to the baseline, the performance improvement on the ASVspoof 2021 LA evaluation set is 31\%. Competitive results are maintained on the ASVspoof 2021 DF dataset.

Rest of the paper is organized as follows: Section 2 describe our method. Experiments, results, and discussions are reported in Section 3. Finally, we conclude the paper in Section 4.

\begin{figure}[th]
  \includegraphics[width=\linewidth]{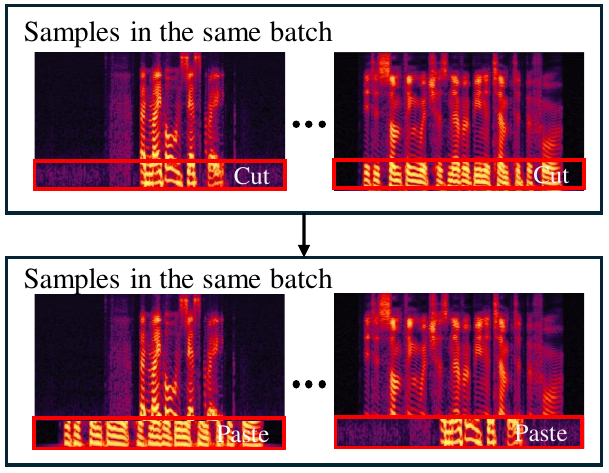}
  \caption{The illustration of the cut and paste operation among the samples in the same batch}
  \label{fig:cut_paste}
\end{figure}

\section{Proposed Method}
In this section, we show the detailed method of FKD, and its overall architecture is shown in Figure 1. First we introduce the process of Freqmix, and then we introduce the specific distillation strategy.

\subsection{Freqmix Data Augmentation}

Inspired by \cite{kim21c_interspeech} \cite{zhong2020random} and our previous work \cite{dong2023multi}, we further develop Specmix and introduc the Freqmix method, a frequency domain DA method. The cut and paste operation is employed on spectrogram of samples within the same batch. Like Specmix, it does not introduce external data. In the cut and paste operation, the content of the identical frequency range (\(f_0, f_0+f\)) in each sample is cut out, shuffled and then pasted back into the corresponding frequency range (\(f_0, f_0+f\)) of the original samples, following the order established during the shuffle. The cut and paste operation is shown in Figure ~\ref{fig:cut_paste}.


To ensure the student model effectively restores the original information and learns the enhancement effects of spectrogram masking, the teacher model must have both the raw audio information and the information after applying the DA method of spectrogram masking. Freqmix leaves part of the teacher model's input unchanged, while applying spectrogram masking to the other part of the data. This allows the student model to learn the original information, compensating for the information distortion introduced by Rawboost, the time-domain DA method, and also combines the DA effects of spectral masking with Rawboost to further improve the model's generalization ability. Therefore, before applying the spectral masking operation to the input of the teacher model, a decision must be made on whether or not to perform cut and paste operation. First, a random value, \(p\), is generated. If \(p > 0.5\), the input data for the teacher model is subjected to the cut and paste operation. After processing through the teacher model, the effects of spectrogram masking can be transferred to the student model. This enables the data of the student model to integrate the enhancement effects of Rawboost and spectrogram masking, the generalization ability of the student model is improved. On the other hand, when \(p < 0.5\), the input data to the teacher model is the original data. This portion of the data is utilized to correct potential information loss in the student model caused by Rawboost, artefacts can be restored, the information extraction ability of student model is enhanced.

\subsection{Knowledge Distillation}

MPIF-Res2Net\cite{dong2023multi}, within the same channel group, merges information from different receptive fields through channel attention. This addresses the issue of information redundancy caused by a single receptive field in Res2Net, mitigating the problem of masking useful information. As shown in Figure ~\ref{fig:speech_production}, the teacher model and the student model constitute the two parts of our FKD method. The input data of the teacher model undergoes DA with Freqmix, in order to cope with the FSD task in real communication scenarios, the input of the student model is enhanced through Rawboost. The student model learns only the relevant knowledge of the output features of each layer from the teacher model, ignoring the prediction results of the teacher model. The purpose is to enable the student model's feature, $SFeat_i$, to fully capture the features of the teacher model's feature, $TFeat_i$, independently of the teacher model's prediction results. This is advantageous for more effective information recovery and the integration of different DA effects. The feature loss function between the student model and the teacher model uses the Mean Squared Error (MSE) function. The MSE function is highly sensitive to the difference between two inputs, making it an appropriate choice for a feature loss function. The calculation method for the feature loss, denoted as $L_{feat}$, is as follows:
\begin{align}
  Loss_{feat} &= \sum_{i=1}^4 MSE(TFeat_i, SFeat_i)
\end{align} Here, \(TFeat_i\) and \(SFeat_i\) respectively denote the outputs of the \(ith\) feature layer for the teacher and student model. In the end, the feature loss \(Loss_{feat}\) is obtained.

For the loss function $Loss_{hard}$ between the student model's predictions and labels, we employ the $A\_softmax$ \cite{liu2017sphereface}:
\begin{align}
  Loss_{hard} &= A\_softmax(predict, label)
\end{align} Here, \(predict\) represents the prediction outcome of the student model, and \(label\) denotes the label of samples in the dataset. \(A\_softmax\) corresponds to our loss function.

Finally, the ultimate loss function $loss$ is obtained by taking the weighted sum of $Loss_{feat}$ and $Loss_{hard}$:
\begin{align}
loss &= \alpha Loss_{feat} + \beta Loss_{hard}
\end{align}Where $\alpha$ and $\beta$ are the weight of $Loss_{feat}$ and $Loss_{hard}$.

The training of the teacher model and the student model are independent. During the distillation stage, the parameters of the teacher model are frozen and do not participate in the update. In the pre-training stage, the teacher model only uses Rawboost for DA. The purpose of this is to be consistent with the student model. MPIF-Res2Net is selected as the classification model. According to our experience, the model based Res2Net can adapt well to the time-frequency features.

\section{Experiments and Results}

\subsection{Datasets}
Two datasets are used in our experiments. The progress subset of the ASVspoof 2021 LA dataset comprises 1,676 genuine speech samples and 14,788 synthesized speech samples. The evaluation subset of ASVspoof 2021 LA dataset has 14,816 bona fide samples and 133,360 spoof utterances, all these samples are transmitted through authentic telephony systems. ASVspoof 2021 LA task aims to design the spoofing counter meatures to enhance the capability of models to detect fake speech under various unknown channel variations. The primary metric for this task are Equal Error Rate (EER) and the minimum tandem detection cost function (t-DCF). The evaluation subset in the ASVspoof 2021 DF dataset consists of a greater number of samples. These samples have been subjected to a sequence of encoding and decoding steps, introducing distortions associated with the codec devices. The evaluation process specifically focuses on the EER as the primary metric.

\subsection{Experimental Setup}

Training a model on a clean dataset and achieving robust performance on an evaluation set with communication interference poses a significant challenge. During the pre-training stage, we employ Rawboost augmentation to the input of teacher model, introducing noise to the original waveform. The noise types include impulsive signal-dependent (ISD) additive noise and stationary signal-independent (SSI) additive noise.

The network takes subbands of the log power spectrum (LPS) in the range of 0-45Hz as input. In the Freqmix DA method, the maximum frequency span $f$ during frequency masking does not exceed 10Hz, the value for deciding whether to apply masking enhancement to the input of the teacher model is set to 0.5. In the short-time Fourier transform, we utilize the Blackman window function, with a window length of 1728 and a hop length of 130. Inspired by \cite{fan2024spatial}, the frequency band of 0-45 dimension is retained, truncated, and flipped before concatenation, resulting in a fixed frame length of 600. The final feature input to the network is a 45 $\times$ 600 F0 sub-band feature map. We employ the MPIF-Res2Net network as the classifier, with Adam used as the optimizer, a batch size of 32, and a learning rate of 0.0001. The random seed is initialized to 1. In the final loss calculation, the weights $\alpha$ and $\beta$ for feature loss and hard loss are set to 0.2 and 0.8, respectively.

\begin{table}[th]
\centering
    \caption{Results of ablation study of the proposed FKD system on the progress and evaluation subset of ASVspoof 2021 LA dataset. Tea means the teacher model, our baseline, before KD; Stu denotes the student model, after KD; The "F" and "R" in parentheses indicate applying the Freqmix and Rawboost DA method to the model's input, respectively. "C" means the clean data. Our experimental results are highlighted in bold. }
    \setlength{\tabcolsep}{4pt}
\begin{tabular}{@{}cccccccc@{}}

\toprule
\multirow{2}{*}{\textbf{System}} & \multicolumn{3}{c}{\textbf{Evaluation set}} &  & \multicolumn{3}{c}{\textbf{Progress set}} \\ \cmidrule(lr){2-4} \cmidrule(l){6-8} 
                        & \textbf{t-DCF}        &       & \textbf{EER (\%)}       &  & \textbf{t-DCF}        &        & \textbf{EER (\%)}     \\ \midrule
Tea(R)                  & 0.2910       &       & 4.19      &  & 0.2860       &        & 4.77        \\
Tea(F+R)                 & 0.2547       &       & 3.31      &  & 0.2250       &        & 3.04       \\
\midrule
Tea(C)\_Stu(R)          & 0.2520       &       & 3.10      &  & 0.2236       &        & 2.92       \\
\textbf{FKD (ours)}          & \textbf{0.2460}       &       & \textbf{2.88}      &  & \textbf{0.2184}       &        & \textbf{2.75}       \\ \bottomrule
\end{tabular}
\end{table}

\subsection{Results}
\subsubsection{Ablation Study}

Table 1 presents the results of various experiments on the progress and evaluation subsets of ASVspoof 2021 LA dataset. Tea(R) serves as the baseline, employing Rawboost directly on MPIF-Res2Net inputs. The approach for Tea(F+R) involves simultaneously applying Freqmix and Rawboost for MPIF-Res2Net inputs. Experiments Tea(C)\_Stu(R) and FKD show the results of the student model after distillation. In Tea(C)\_Stu(R), the teacher model employs the original dataset, while the student model is enhanced with Rawboost. In FKD, the teacher model's input incorporates Freqmix, while the student model's input uses Rawboost. Tea means the teacher model, Stu means the student model. Notably, the Tea(R) is our baseline, the inputs were augmented by Rawboost, lacking original speech restoration, attains an EER of 4.19$\%$ and a t-DCF of 0.2910 on the evaluation subset. The experiment Tea(C)\_Stu(R) shows when the teacher model uses the original data to perform information compensation on the Rawboost-enhanced input of the student model, the obtained EER and t-DCF are 3.10$\%$ and 0.2520 respectively. This confirms that DA may cause information loss and using original data for information recovery is a viable approach to adjusting the extent of DA, thereby enhancing model performance. On the other hand, in the Tea(F+R) experiment where Freqmix and Rawboost data enhancement were applied to the MPIF-Res2Net model at the same time, the EER was 3.31$\%$ and the t-DCF was 0.2547, which seems to be a good result. However, in the FKD experiment, where Freqmix data augmentation was used as input to the teacher model, the student model achieved EER and t-DCF results of 2.88$\%$ and 0.2460, respectively. We attribute this improvement to the fusion of DA methods and the compensation of information in the student model input. Specifically, Freqmix retains some information in the original data to compensate for the information loss caused by Rawboost enhancement. In addition, the teacher model imparts the masking enhancement effect in the two-dimensional spectrogram to the data in the student model, so that the masking enhancement effect of Freqmix and communication enhancement effect of Rawboost are integrated in the student model. Therefore, the performance of the model is further improved.

\begin{table}[th]
    \caption{Results Comparison with Fusion Systems on the Performance
    of progress and evaluation subset of ASVspoof 2021 LA Dataset. Our experimental results are highlighted in bold.}
    \centering
    \setlength{\tabcolsep}{2pt}
\begin{tabular}{@{}cccccccc@{}}
\toprule
\multirow{2}{*}{\textbf{System}} & \multicolumn{3}{c}{\textbf{Evaluation set}}   & \textbf{} & \multicolumn{3}{c}{\textbf{Progress set}} \\ \cmidrule(lr){2-4} \cmidrule(l){6-8} 
                                 & \textbf{t-DCF}  & \textbf{} & \textbf{EER (\%)}  & \textbf{} & \textbf{t-DCF}  & \textbf{} & \textbf{EER (\%)}  \\ \midrule
T06 \cite{liu2023asvspoof}                                                   & 0.2853          &           & 5.66     & & 0.2476          &           & 5.61     \\
Fusioin systems \cite{cohen2022study}                                        & 0.2882          &           & 4.66     & & --              &           & --     \\
T36 \cite{liu2023asvspoof}                                                   & 0.2531          &           & 3.10      &  & 0.2373          &           & 3.69    \\
T35 \cite{liu2023asvspoof}                                                  & 0.2480          &           & 2.77   &     & 0.2115          &           & 2.61  \\
T23 \cite{liu2023asvspoof}                                                  & 0.2176          &           & 1.32          &        & 0.1816          &           & 0.89\\

\textbf{FKD (ours)}                  \textbf{} & \textbf{0.2460} & \textbf{} & \textbf{2.88} & & \textbf{0.2184} & \textbf{} & \textbf{2.75}\\ \bottomrule
\end{tabular}
\end{table}

\begin{table}[th]
\centering
    \caption{Results Comparison with Fusion Systems on the Performance
    of progress and evaluation subset of ASVspoof2021 LA Dataset. Our experimental results are highlighted in bold.}
\begin{tabular}{@{}cccc@{}}
\toprule
\textbf{System}   & \textbf{min t-DCF} & \textbf{}            & \textbf{EER(\%)} \\ \midrule 
RawGAT-ST \cite{tak2021end}         & 0.3782             & \multicolumn{1}{l}{} & 6.92             \\
M-GMM-MobileNet(C) \cite{wen2022multi}     &0.3231            &      &6.80 \\
RawNet2 \cite{tak2021end}           & 0.3099             &                      & 5.31             \\
AASIST \cite{jung2022aasist}            & 0.3398             &                      & 5.59             \\
DFIM \cite{huang2023discriminative}              & 0.2601             &                      & 3.05             \\

\textbf{FKD (ours)} & \textbf{0.2460}    & \textbf{}            & \textbf{2.88}    \\ \bottomrule
\end{tabular}
\end{table}

\subsubsection{Performance Comparison With Other Systems}
Table 2 shows the results of the FKD method and other different fusion models on the eval subset and progress subset of ASVspoof 2021 LA dataset. The comparison shows that the performance of our method even exceeds some fusion systems. Although the results of T23 and T35 are better than our model, their fusion strategies are quite complex. For example, the T23 fusion system consists of multiple subsystems extracting different spectral features from various codec-augmented data to train different classifiers, and finally performs weighted average score fusion. Additionally, the t-DCF value of 0.2480 on the eval subset for T35 is slightly higher than the t-DFC value of our FKD system. This indicates that the reliability of the fusion system T35 is slightly lower than our approach when working in tandem with the ASV system.

Table 3 shows the effect of our method on the ASVspoof 2021 LA evaluation subset compared with the FKD system and other single systems. It can be seen that our method has good performance compared with other single systems. Compared to the best-known experimental results to date, our method has shown improvements in both EER and t-DCF. This indicates that our method yields more accurate classification results and is more reliable when used in tandem with an ASV system.

\begin{table}[th]
\centering
\caption{Results Comparison with single Systems on the Performance
     of ASVspoof2021 DF Dataset. Our experimental results are highlighted in bold.}
\begin{tabular}{cc}
\hline
\textbf{System}              & \textbf{21DF EER(\%)} \\ \hline
ARawNet2 \cite{li23h_interspeech}                     & 19.03                 \\
T06 \cite{liu2023asvspoof}                         & 19.01                  \\
T22 \cite{liu2023asvspoof} & 19.22                 \\
CQT-LCNN(D3) \cite{das2021known}                    &18.31 \\
T08 \cite{liu2023asvspoof}                   & 18.30  \\


ResNet-S-LDE \cite{chen2021pindrop}                  & 17.25                  \\

\textbf{FKD (ours)}            & \textbf{17.16}                 \\ \hline
\end{tabular}
\end{table}

Table 4 presents a comparison of the results between FKD and other methods. It can be observed that our approach still maintains competitive results on the DF dataset.


\section{Conclusion}

In this article we propose an approach that combines the DA and KD methods. First, we introduce the Freqmix DA method. This method divides the input of the teacher model into two parts and imparts two types of knowledge to the teacher model: original information and the DA effect of spectrogram masking. In the KD method, the teacher model imparts both types of knowledge to the student model. Specifically, the original information in the teacher model helps to restore some artefacts in the student model caused by Rawboost, thereby improving the information extraction ability of the student model. At the same time, the DA effect of spectrogram masking in the teacher model is transferred to the student model and merges with the Rawboost in the student model, further improving the generalisation ability of the model. Our experimental results surpass the best results of all individual systems, demonstrating the effectiveness of combining DA and KD methods in the FSD task under telephony scenarios.

\section{Acknowledgements}
This work is supported by the {STI 2030—Major Projects (No. 2021ZD0201500)}, the National Natural Science Foundation of China (NSFC) (No.62201002, No.62322120), Distinguished Youth Foundation of Anhui Scientific Committee (No. 2208085J05), Special Fund for Key Program of Science and Technology of Anhui Province (No. 202203a07020008), Open Fund of Key Laboratory of Flight Techniques and Flight Safety, CACC (No, FZ2022KF15).

\bibliographystyle{IEEEtran}
\bibliography{mybib}

\end{document}